\documentclass[12pt]{article}
\usepackage{scicite}
\usepackage{times}
\usepackage{graphicx} 
\usepackage{rotating}
\usepackage{svg}
\usepackage{lscape}
\usepackage{array} 
\usepackage{amsmath}
\usepackage{url}
\usepackage{hyperref}
\usepackage{adjustbox}

\newenvironment{sciabstract}{%
\begin{quote} \bf}
{\end{quote}}

\title{Computational analysis of US Congressional speeches reveals a shift from evidence to intuition} 
\author
{Segun T. Aroyehun,$^{1\ast}$ Almog Simchon,$^{2}$ Fabio Carrella,$^{3}$ \\ 
Jana Lasser,$^{5,6}$ Stephan Lewandowsky,$^{3,4}$ David Garcia$^{1,6}$\\
\\
\normalsize{$^{1}$University of Konstanz, Germany}\\
\normalsize{$^{2}$Ben-Gurion University of the Negev, Israel}\\
\normalsize{$^{3}$University of Bristol, UK}\\
\normalsize{$^{4}$University of Potsdam, Germany}\\
\normalsize{$^{5}$Graz University of Technology, Austria}\\
\normalsize{$^{6}$Complexity Science Hub, Vienna, Austria}\\
\\
\normalsize{$^\ast$To whom correspondence should be addressed; E-mail:   segun.aroyehun@uni-konstanz.de}
}

\date{}

\begin{document} 

\baselineskip24pt

\maketitle 

\begin{sciabstract}
Pursuit of honest and truthful decision-making is crucial for governance and accountability in democracies. However, people sometimes take different perspectives of what it means to be honest and how to pursue truthfulness. Here we explore a continuum of perspectives from evidence-based reasoning, rooted in ascertainable facts and data, at one end, to intuitive decisions that are driven by feelings and subjective interpretations, at the other. We analyze the linguistic traces of those contrasting perspectives in Congressional speeches from 1879 to 2022. We find that evidence-based language has continued to decline since the mid-1970s, together with a decline in legislative productivity. The decline was accompanied by increasing partisan polarization in Congress and rising income inequality in society. Results highlight the importance of evidence-based language in political decision-making.
\end{sciabstract}

\section*{Main text}

Honesty and truthfulness underpin accountability, transparency, and informed decision-making in democratic societies. A collective commitment to truth cultivates discourse grounded in empirical evidence and fosters social cohesion through a shared understanding of reality \cite{higgins2021shared}.
In many democracies, there is currently much concern about ``truth decay'' \cite{kavanagh2018truth}: the blurring of the boundary between fact and fiction
\cite{lewandowsky2017beyond}, not only fuelling polarization but also undermining public trust in institutions \cite{bennett2018disinformation, lewandowsky2017beyond}. 

We adopt a framework that distinguishes two rhetorical approaches with which politicians can express
their pursuit of truth
\cite{garrett2017epistemic, lewandowsky2020willful, cooper2023honest, lewandowsky2024liars}. One approach, which we call
evidence-based, pursues truth by relying on evidence, facts, data, and other elements of external reality. An alternative
approach, called intuition-based, pursues truth by relying on 
feelings, instincts, personal values, and other elements drawn mainly from a person's internal
experiences.
Productive democratic discourse balances between evidence-based and intuition-based
conceptions of truth. While evidence-based discourse provides a foundation for ``reasoned'' debate, 
intuition contributes emotional and experiential dimensions that can be critical for exploring and resolving societal issues. 
However, although the mix of evidence-based and intuition-based
pathways to truth ranges along a continuum, exclusive reliance on intuition
may prevent productive political debate because evidence and data can no longer adjudicate between competing political positions and eventually lead to an agreement. 
Here we examine these developments by analyzing the basic conceptions of truth 
that politicians deploy in political speech. 
We are not concerned with the truth value of individual assertions but with how the 
pursuit of truth is reflected in political rhetoric.  

We apply computational text analysis  \cite{NIPS2013_9aa42b31, garten2018dictionaries}
to measure the relative prevalence of evidence-based and intuition-based language in 145 years of speeches on the floor of the U.S. Congress. The conceptions of truth employed in Congressional rhetoric are relevant to various measures of political and societal welfare.
We analyze Congressional rhetoric in relation to two likely drivers of democratic backsliding \cite{Szostak24}:  partisan polarization and income inequality.
Polarization, characterized by growing ideological divisions and partisan animosity, undermines constructive dialogue, hampers compromise, and erodes trust in political institutions, ultimately weakening democratic processes \cite{graham2020democracy,finkel2020political}.
Previous research underscores the link between political polarization and language use, highlighting the influence of ideological divisions on communication patterns and political behavior \cite{gentzkow2010drives,jensen2012political}.
Economic inequality also exerts negative effects on various individual and social outcomes \cite{polacko2021causes}. For example, individuals in environments characterized by high inequality 
tend to project individualistic norms onto society \cite{sanchez2019economic}. This fosters greater competition and reduces cooperation, which in turn may damage democracy \cite{Szostak24}. 
Polarization can play a role in increasing inequality through lower congressional productivity \cite{mccarty2016polarized}, which could be affected by a shift from evidence-based language to intuition-based language in Congressional rhetoric. This motivates our analysis of Congressional rhetoric in Congressional productivity, as assessed through the quantity and quality of enacted laws over time \cite{grant2008legislative,libgober2024comprehensive}.

\paragraph{\textit{Measure of evidence-based and intuition-based language}}\mbox{}\\
Our analysis involves 8 million Congressional speech transcripts between 1879 and 2022. 
Details on pre-processing of the corpus can be found in the Supplementary materials.
We measure the relative salience of evidence-based language over intuition-based language as the Evidence-Minus-Intuition (EMI) score, building on a text analysis approach that combines dictionaries with word embeddings to represent documents and concepts \cite{garten2018dictionaries} as used in previous work on political communication \cite{gennaro2022emotion,lasser2023alternative}. We constructed dictionaries to capture evidence-based and intuition-based language styles that underlie the two conceptions of truth (e.g. ``fact'' and ``proof'' in the evidence-based dictionary and ``guess'' and ``believe'' in the intuition-based dictionary, see Supplementary materials for the full dictionaries).
We adopt the approach for construction and validation of dictionaries
used in \cite{lasser2023alternative} (see Supplementary materials for details).
Our final dictionaries consist of 49 keywords for evidence-based language and 35 keywords for intuition-based language (see Tables \ref{tab:evidence_keywords} and \ref{tab:intuition_keywords} in the Supplementary materials).
We employ a Word2Vec embeddings model \cite{NIPS2013_9aa42b31} that we train on the Congressional speeches. This approach converts each conception of truth into a vector representation by averaging the embeddings of the corresponding dictionary keywords. 
Similarly, the target text is represented as the average word embeddings of content words. 
We quantify the EMI score as the difference between cosine similarities of the text
being analyzed and the two dictionaries. A positive EMI score indicates a higher prevalence of evidence-based language (see Table \ref{tab:pos_emi} for examples), whereas a negative score suggests reliance on intuition-based language (refer to Table \ref{tab:neg_emi} for examples). 
See Supplementary materials for further details, including validation of the EMI score against human ratings.

\begin{figure}
    \centering
    \includegraphics[width=\textwidth, height=0.85\textheight, keepaspectratio]{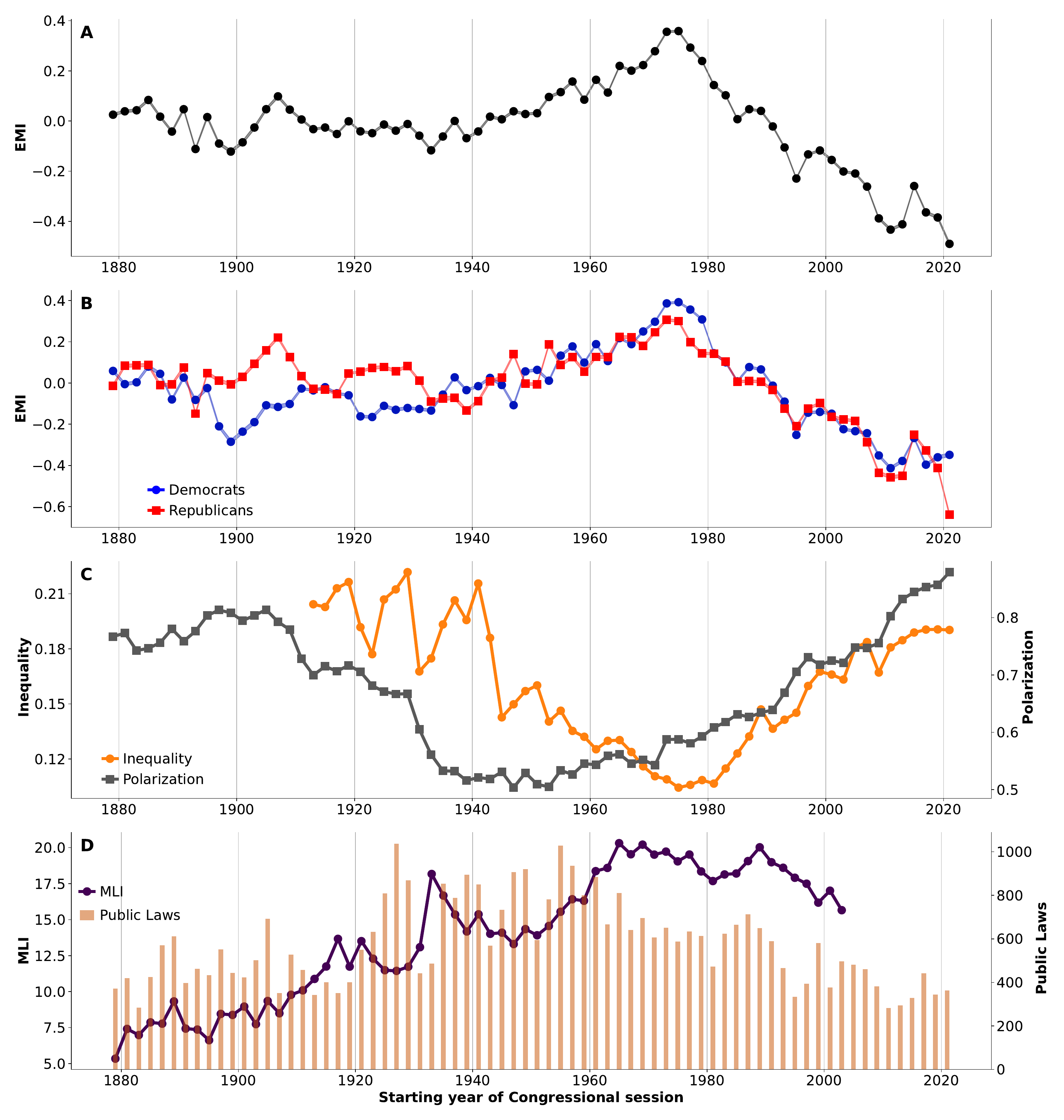}
    \caption{Time series of Evidence-Minus-Intuition (EMI) score in each congressional session between 1879 and 2022 (A), EMI scores separated by party (B), congressional polarization and inequality (C) and congressional productivity, measured as the Major Legislation Index (MLI) and the number of public laws passed by each session (D). We compute bootstrapping 95\% CIs for EMI with 10000 samples, which may appear too small to be visible due to the large sample size. }
    \label{fig:trend_panel}
\end{figure}

\paragraph{\textit{Trend of EMI over time, by party, and across chambers}}\mbox{}\\
Figure \ref{fig:trend_panel}A shows the trend of EMI score over time, reflecting the relative prevalence of evidence-based language. EMI was high and relatively stable from 1875 through
the early part of the 20$^{th}$ century. Subsequently, an upward trend from the 1940s 
culminated in a peak in the mid-1970s. Since then, evidence-based language has been on the decline. 

However, significant dips in EMI also occurred during the early, stable period: 
The 56th Congress (from 1899 to 1901) has the historically lowest EMI score prior to the 1970s, closely followed by the 73rd Congress (from 1933 to 1935). 
These two periods align with significant historical events. In the 1890s, the U.S. experienced the Gilded age,
marked by rapid industrialization and economic growth, but also social unrest and increasing economic inequality. The 1930s were marked by the Great Depression during which the country faced high unemployment, widespread poverty, and social upheaval. 
These economic and social upheavals likely influenced the language used in Congress during these periods. The profound impact of these events might have led to a greater emphasis on intuition-based language,
consonant with previous research that has documented shifts in language use 
among individuals facing stressful situations \cite{pennebaker2003psychological} as well as among political leaders confronted with crises \cite{wallace1993political,pennebaker2002language}.
An examination of a sample of speeches with low EMI score in specific periods shows a tendency 
to focus on the crisis of the time (see Table \ref{tab:sample_periods} for illustrative examples).

Focusing on the period past 1970, one striking observation is that the level of EMI has
recently fallen to its historical minimum, following a decreasing linear trend that started in the peak session of 1975-1976 ($b=-0.032$,  $p<0.001$,  $R^2= 0.927$).
Figure \ref{fig:trend_panel}B illustrates the temporal trend of the EMI score for Democrats and Republicans separately. There is a strong positive correlation between the EMI scores for both parties ($r = 0.778$, $95\%CI = [0.666, 0.855]$, $p<0.005$).
We observe some divergence between parties in the early periods. However, since the mid-70s, both parties have moved largely in the same downward direction in their rhetoric. The same pattern holds for both parties across the House and Senate (Figure \ref{fig:emi_party_chamber} in the Supplementary materials). It is however noticeable that the EMI of Republicans dropped substantially, and
more steeply than for Democrats,
in the last session (2021-2022). A Mann-Whitney test shows that the difference in median EMI score ($-0.435$ for democrats and $-0.753$ for republicans) is significant ($p<0.001$). 

Turning to the implications of the observed trends, 
Figure \ref{fig:trend_panel}C 
shows partisan polarization in Congress over time, measured as the difference between the first dimension of DW-NOMINATE 
scores \cite{poole2011ideology, lewis2021voteview} for the two major parties averaged across the House and Senate. 
Figure \ref{fig:trend_panel}C also includes the trend of income inequality \cite{alvaredo2016distributional} using the share of pre-tax income of the top 1\% of the population (source: \url{https://wid.world/}). The recent decline of EMI is accompanied by a corresponding upward trend in partisan polarization in Congress and rising income inequality in society, which is statistically supported as follows.

\paragraph{\textit{EMI and polarization}}\mbox{}\\
EMI and polarization are negatively cross-correlated ($r=-0.615$, $95\%CI = [-0.741,-0.447]$, $p<0.005$) and a lagged correlation analysis shows that lag zero has the highest correlation 
(see Figure~\ref{fig:crosscorr}A). Figure~\ref{fig:pol_emi_scatter} also depicts the relationship. When included in lagged regression models, EMI does not explain a significant amount of the empirical variance of polarization, but polarization has a significant coefficient in the EMI model (Table \ref{tab:EMI-Pol}).

\begin{figure}
    \centering
    \includegraphics[width=\textwidth, height=0.85\textheight, keepaspectratio]{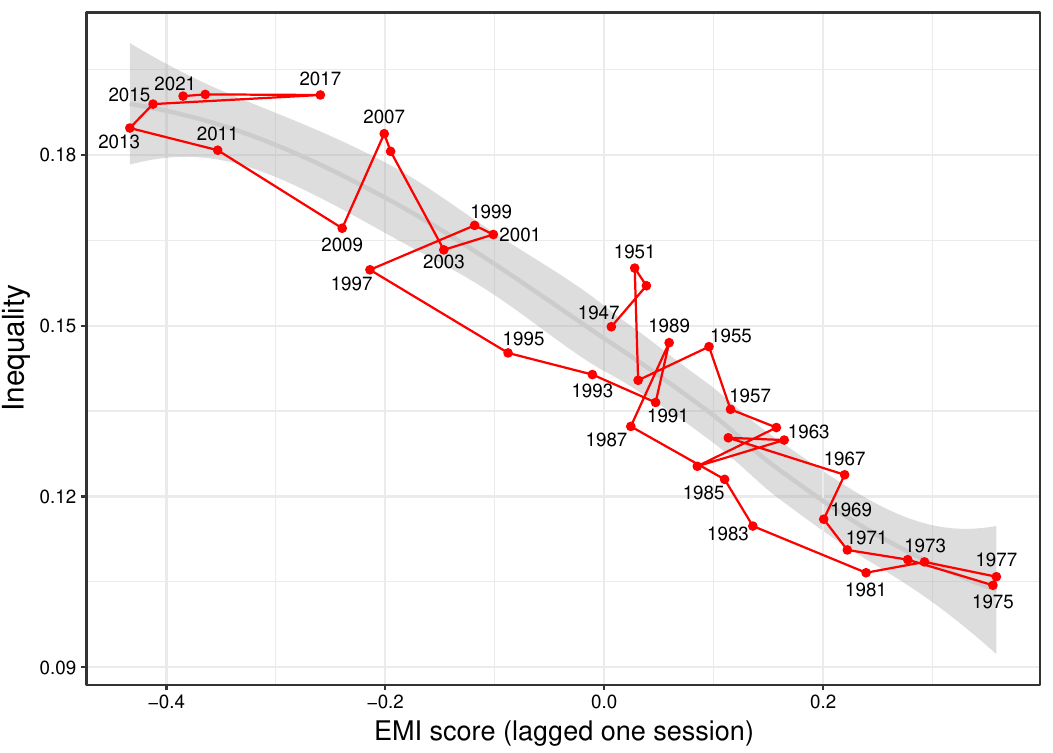}
    \caption{Inequality measured as the share of income of the top 1
\% versus the EMI score in the previous legislature. The shaded area shows a LOESS fit and labels indicate the year corresponding to the inequality measurement.}
    \label{fig:EMI-Ineq}
\end{figure}

\paragraph{\textit{EMI and income inequality}} \mbox{}\\
EMI values are informative of future inequality. Figure~\ref{fig:EMI-Ineq} shows the historical values of inequality as a function of EMI in the previous session, i.e., the previous two years ($r = -0.948$, $95\%CI = [-0.973, -0.902]$, $p<0.001$). A lagged correlation analysis shows that the 
strongest correlations appear when EMI precedes inequality (see Figure \ref{fig:crosscorr}B).

This is buttressed by a lagged regression model including the level of inequality, polarization, and EMI from the previous session, as well as their interaction. The results of that fit are shown in 
Table~\ref{tab:EMI-Ineq} in comparison with an autoregressive model, 
revealing a negative coefficient of EMI with inequality two years later. The interaction with polarization is not significant and weak enough for the slope of EMI to stay negative (see Figure~\ref{fig:interactions}).
These regression results are robust to other specifications of the analysis, for example when using the Gini index instead of the top-1\% share of income 
(restricted to the time since full income data became available), 
when using all available data since 1912, and when considering a longer 
lag for polarization (see Table~\ref{tab:EMI-Ineq-Ext}).

\paragraph{\textit{Relationship between EMI and Congressional productivity}}\mbox{}\\
Evidence-based language can be a tool to identify 
factual constraints for Congress to formulate legislation, which often
requires some form of bipartisan agreement. 
We examine the relationship between EMI and Congressional productivity as measured by three indicators. First is the Major Legislation Index (MLI)  \cite{grant2008legislative} which measures the productivity of Congress in terms of important legislation. Second is the Legislative Productivity Index (LPI) \cite{grant2008legislative}, which combines assessments of important legislation and number of laws enacted. Third is the count of the number of laws passed by each session of Congress \cite{libgober2024comprehensive} without considering their significance. Previous research analyzes congressional productivity as a function of polarization, party composition in the legislature and executive branch \cite{grant2008legislative}, and public mood towards more regulation as measured in surveys \cite{stimson2018public}. From these indicators, polarization and public mood toward regulation are the most important predictors, explaining a significant amount of the variance of productivity over time \cite{grant2008legislative}.

\begin{figure}
    \centering
    \includegraphics[width=\textwidth, height=0.85\textheight, keepaspectratio]{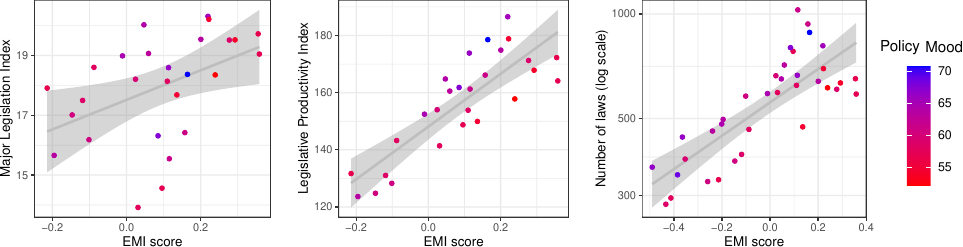}
    \caption{EMI score versus Congressional productivity measured as MLI (left), LPI (center), and log-transformed number of laws (right). Points are colored according to public mood towards regulation during the legislative period and gray lines and shaded areas show linear regression models of each productivity variable as a function of EMI alone.}
    \label{fig:EMI-Prod}
\end{figure}

Figure~\ref{fig:EMI-Prod} shows the relationship between all three Congressional productivity metrics and EMI measured in the same session. All three cases have positive and significant correlations (MLI: $r=0.454$, $95\%CI=[0.09, 0.711]$, $p<0.05$; LPI: $r=0.836$, $95\%CI=[0.667, 0.923]$, $p<0.001$; log-transformed number of laws: $r=0.796$, $95\%CI=[0.633, 0.891]$, $p<0.001$). However, polarization and public mood about regulation play an important role in Congressional productivity, which is shown by the color of plotting symbols
in Figure \ref{fig:EMI-Prod}. Points representing high public mood (blue) tend to 
lie above the regression line, and points with low public mood (red) tend to lie below. 
For that reason, we fitted the base models of \cite{grant2008legislative} and tested if adding the EMI of a session has a positive effect on legislative productivity indices. Results are shown in 
Table~\ref{tab:Prod}, revealing that, after controlling for known effects
in productivity and for an interaction between polarization and EMI, the coefficient of EMI 
is positive and significant for MLI and LPI, and positive for the number of laws but 
only at the $0.1$ level.  
We see this as an indication that EMI plays a role in Congressional productivity, with the effect being more salient when considering major legislation in comparison to minor laws where parliamentary debate might not play a bigger role.

\paragraph{\textit{Discussion and conclusion}}\mbox{}\\
We introduce an approach for quantifying the 
conception of truth that members of Congress embrace and deploy in their rhetoric. 
Using embedded dictionaries in conjunction with embedding of Congressional speeches, we calculate and validate the Evidence-Minus-Intuition (EMI) score from transcripts of Congressional 
speeches spanning the years 1879 to 2022. The EMI score reflects
the prevalence of evidence-based language when positive and intuition-based language when negative. 
We study the temporal trends of the EMI score and investigate its relationships 
with measures of polarization and inequality as well as Congressional productivity. 

We find that EMI shows a pattern of relative stability until the 1940s, which is followed by 
a clear upward trajectory that reached a maximum in the 1970s. Since then, EMI trends downward, indicating a decline in the prevalence of evidence-based language for both parties.
The degree of synchronization in the linguistic styles employed by both Democrats and Republicans during this period points to their alignment around messaging strategies \cite{neiman2016speaking}.

We relate the decline in EMI to three outcome variables that are indicative of democratic
health and find a concerning association in all cases: a decline in evidence-based language 
is associated with increasing polarization and increasing income inequality 
but decreased congressional productivity. The temporal sequence of those trends differs between 
variables. 
For polarization, the association with EMI is greatest at lags zero and polarization is found
to explain EMI, but not vice versa, suggesting that polarization is an immediate driver
of politicians' rhetoric. 
By contrast, EMI precedes shifts in income inequality, such that a stronger emphasis on evidence-based reasoning 
is followed by reduced income inequality whereas greater reliance on intuition seems to
perpetuate existing social disparities. This finding aligns with existing research on language and social inequality \cite{Augoustinos2019}, which underscores how language patterns have consequences for understanding social issues and may either promote or inhibit necessary changes.
Intuition-based language can explain the relationship between polarization and inequality, as it hinders legislation that addresses income inequality through redistribution \cite{mccarty2016polarized}.

Finally, the impact of evidence-based language on Congressional productivity is again
contemporaneous.
In the Habermasian view of communicative action \cite{habermas1984theory}, evidence-based language serves as the foundation for ``reasoned'' debates and can steer discussions away from personal and political hostilities. 
In this communicative process, evidence-based language serves as a tool to establish a
shared understanding of the state of the world and contributes to the formulation
of well-informed decisions. 
The positive correlation that we observe in our study between the EMI score and legislative productivity (in terms of quality and quantity) is in line with this viewpoint. 

The observed patterns in Congressional language are the result of a complex interplay of various factors, some of which are unique to the political and societal context of the United States. One contributing factor to these patterns is the control exerted by party leadership over who speaks on the Congressional floor \cite{morris2001reexamining}, potentially shaping the content and tone of speeches. This control mechanism is likely to influence the language used by Congressional members in aligning with the strategic objectives of their party. 
In addition to the influence exerted by party leaders, members of Congress may find themselves compelled to cater to their base, encompassing constituents, donors, and lobbyists, particularly in a highly polarized environment driven by partisanship \cite{wilson2020polarization}.

Furthermore, the impact of media on politicians, particularly their adoption of media logic \cite{altheide2004media}, introduces an additional dimension to the nature of political representation. This influence could be amplified by the live coverage of proceedings through the C-SPAN (first introduced in the House in 1979 and then in the Senate in 1986). In an era characterized by increasing polarization, politicians might find themselves driven to embrace a perpetual campaign style of representation \cite{Esser2013}, transforming Congressional speeches into orchestrated performances aimed at capturing media attention. Consequently, this shift may result in a reduced focus on meaningful intellectual discourse and nuanced policy discussions within the legislative body.
This interpretation meshes well with a recent analysis of the Twitter/X
communications of U.S. Congress members from 2011 to 2022, which similarly differentiated
between evidence-based ``fact-speaking'' and authentic ``belief-speaking'' as alternative
expressions of honesty \cite{lasser2023alternative}.
That study discovered an association between the prevalence of authentic ``belief-speaking'' and a 
decrease in the quality of shared sources in tweets, particularly among Republicans. This suggests a potential link between belief-based language and the dissemination of low-quality information
to the public.

We have highlighted the adverse consequences
of a Congress characterized by diminishing evidence-based language and rising partisan polarization.
The decline in the quality and quantity of legislative output at a time of multiple global
crises should be of concern. 
On a more positive note, understanding the complex relationship between the language of political discourse and partisan polarization points to avenues for interventions focused on fostering more constructive and productive debate. Initiatives such as those promoting collaboration and communication across partisan boundaries \cite{hartman2022interventions} can contribute to rebuilding a more robust democratic discourse.
Ultimately, the challenge lies in having a Congress (and by extension a deliberative public) where truth is valued, polarization is in check, and legislative outcomes reflect the diverse needs of the citizens.

\newpage

\bibliographystyle{Science}

\section*{Acknowledgments}
\textbf{Funding:} SL acknowledges financial support from 
the European Research Council
(ERC Advanced Grant 101020961 PRODEMINFO), the 
Humboldt Foundation through a research award, the 
Volkswagen Foundation (grant ``Reclaiming individual autonomy and democratic discourse online: How to rebalance human and algorithmic decision making''),
and the European Commission (Horizon 2020 grant 101094752 SoMe4Dem).
SL also receives funding from Jigsaw (a technology incubator created by Google) and from UK Research and Innovation through
EU Horizon replacement funding grant number 10049415.
DG is also a beneficiary of the ERC Advanced Grant 101020961 PRODEMINFO
and STA is employed by PRODEMINFO.
DG also received funding from the Deutsche Forschungsgemeinschaft (DFG – German Research Foundation) under Germany's Excellence Strategy – EXC-2035/1 – 390681379.
JL was supported by the Marie Skłodowska-Curie grant No. 101026507. The authors thank Tyler Brown for useful input on the manuscript.\\
\textbf{Author contributions:} SL, DG and STA conceptualised the research. STA collected the data. STA and DG developed the text analysis pipeline. FC performed the construction and validation of the keywords. AS performed the validation of the EMI score. DG and STA performed the statistical analyses. JL provided advice on the statistical analyses and visualization. SL and DG acquired funding and supervised the project. STA , DG, and SL prepared the initial draft of the manuscript. All authors contributed to preparing and editing the final version of the manuscript. \\
\textbf{Competing interests:} Authors have no competing interests.\\
\textbf{Data and materials availability:}
The datasets used in this study are deposited in an Open Science Framework (OSF) repository (\url{https://doi.org/10.17605/OSF.IO/Z6UTW}) \cite{data_repo}. 
The codes used to perform the analyses reported in this paper are available in a Github repository (\url{https://doi.org/10.5281/zenodo.11127530}) \cite{code_repo}. \\ \\

\newpage
\setcounter{page}{1}
\setcounter{figure}{0}
\setcounter{table}{0}

\renewcommand{\thefigure}{S\arabic{figure}}
\renewcommand*\figurename{fig.}
\renewcommand*\tablename{table}
\renewcommand{\thetable}{S\arabic{table}}
\begin{center}
Supplementary materials for \\
\textbf{Computational analysis of US congressional speeches reveals a shift from evidence to intuition} \\
\textit{Segun T. Aroyehun et al.} \\
\textit{Corresponding author: Segun T. Aroyehun, segun.aroyehun@uni-konstanz.de}\\
    
\end{center}
\section*{Materials and methods}
\subsection*{Data pre-processing steps}
We initially rely on the dataset compiled by Gentzkow et al. ~\cite{gentzkow2018congressional} and supplement it with recent data obtained by accessing the Congressional records' website using an automated script \cite{judd2017congressional}. The dataset includes essential metadata such as speaker information (including party) and dates. 
The dataset consists of 14,153,443 speeches spanning the Congressional sessions of 1879 to 2022. To ensure the quality of our dataset, we employ a number of pre-processing steps. First, we remove procedural speeches. Procedural speeches are speeches delivered by members of Congress that mainly deal with the rules and procedures that govern legislative proceedings. These may include discussions on amendments to rules, requests for unanimous consent, or the announcement of votes. 
We train three classifiers, following the methodology outlined by Card et al. \cite{card2022computational}, to identify procedural speeches. We remove procedural speeches by employing a majority vote ensemble of the classifiers.

\begin{figure}[h]
    \centering
    \includegraphics[width=\textwidth, height=0.75\textheight, keepaspectratio]{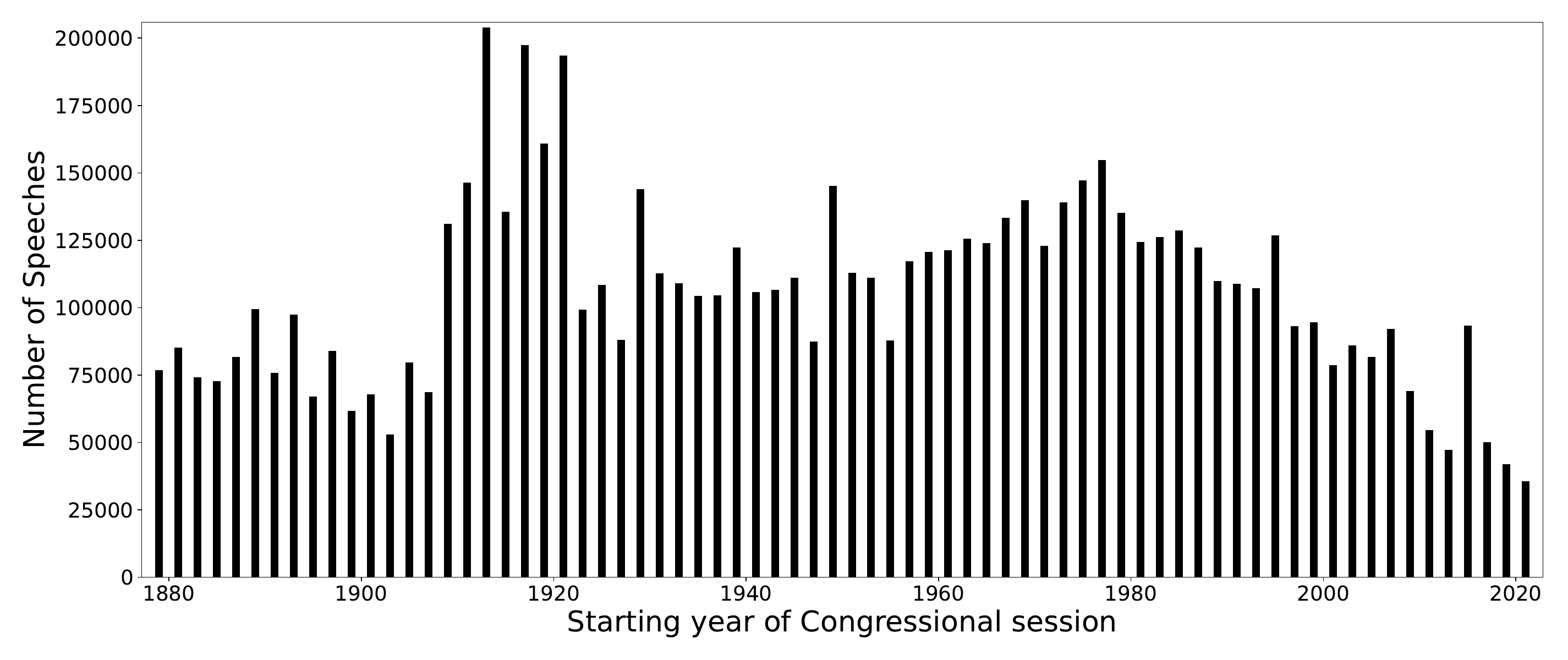}
    \caption{Number of speeches across both House and Senate for Congressional sessions between 1879 and 2022}
    \label{fig:speech_count}
\end{figure}

In general, the Congressional Record is of high quality. However, in the earlier years, it contains some instances of optical character recognition errors that result in unintelligible content (e.g., in the rendering of a table).
To mitigate the potential noise from lengthy speeches that mainly consist of lists of names or numbers, we employ a filtering mechanism. This filter evaluates the ratio of common (top 100) English words (e.g., ``the'', ``and'', ``is'') to the total token length of a speech. We set a threshold of 0.05, ensuring that speeches with substantive content are retained for further analysis. We keep speeches that are attributed to members of the two major parties.
We filter out speeches with fewer than 11 tokens and remove duplicate entries. Our final dataset consists of 8,435,769 speeches with an average length of approximately 199 tokens. Speeches made by Democrats account for 53\% of the dataset and 47\% of speeches are by Republicans. Fig. \ref{fig:speech_count} shows the number of speeches for each congressional session across both chambers (House and Senate) from 1879 to 2022. The number of speeches for each session varies. Nevertheless, there is a substantial amount of speeches available, with at least 35,000 speeches for each session, to enable a reliable analysis.
To facilitate further analysis, we split longer speeches (consisting of more than 150 tokens) into chunks of approximately 150 tokens each. We set a minimum chunk size of 50 tokens, such that a chunk smaller than the minimum size is merged with the immediately preceding chunk.

\newpage
\subsection*{Steps for the construction and validation of EMI score}

\paragraph{List of keywords.}
We constructed dictionaries consisting of words associated with the two conceptions of truth as
reflected in evidence-based and intuition-based language, respectively. We first created a list of seed words and then expanded it computationally \cite{dinatale2024lexpander}. Following the same approach used in \cite{lasser2023alternative}, we then recruited participants on Prolific to rate each keyword on their representativeness on two scales, one for evidence-based and one for 
intuition-based language. We then keep only words rated as statistically more representative for their respective construct than the other. 

Our final dictionaries consist of 49 keywords for evidence-based language and 35 keywords for intuition-based language (see Tables \ref{tab:evidence_keywords} and \ref{tab:intuition_keywords}).
The difference in the number of keywords is not a concern in our approach, since we employ the distributed dictionary representation method \cite{garten2018dictionaries}, which effectively normalizes the impact of varying keyword counts by representing each dictionary with a single vector. This ensures a consistent measure of evidence-based and intuition-based language, enabling meaningful comparisons across both constructs.

\begin{table}[htbp]
    \centering
    \caption{Dictionary for evidence-based language}
    \begin{tabular}{|ccccc|}
         \hline
         \multicolumn{5}{|c|}{Keywords} \\ \hline
         accurate&  exact&  intelligence&  precise& search\\
         analyse&  examination&  investigate&  procedure& show\\
         analysis&  examine&  investigation&  process& statistics\\
         correct&  expert&  knowledge&  proof& study\\
         correction&  explore&  lab&  question& trial\\
         data&  fact&  learn&  read& real\\
         dossier&  find&  logic&  reason& true\\
         education&  findings&  logical&  research& truth\\
         evidence&  information&  method&  science& truthful\\
         evident&  inquiry&  pinpoint&  scientific& \\ \hline
    \end{tabular}
    \label{tab:evidence_keywords}
\end{table}

\begin{table}[htbp]
    \centering
    \caption{Dictionary for intuition-based language}
    \begin{tabular}{|cccc|}
         \hline
         \multicolumn{4}{|c|}{Keywords} \\ \hline
         advice&  doubt&  mislead& suggestion\\
         belief&  fake&  mistaken& suspicion\\
         believe&  fake news&  mistrust& view\\
         bogus&  feeling&  opinion& viewpoint\\
         common sense&  genuine&  perspective& wrong\\
         deceive&  guess&  phony& \\
         deception&  gut&  point of view& \\
         dishonest&  instinct&  propaganda& \\
         dishonesty&  intuition&  sense& \\
         distrust&  lie&  suggest& \\ \hline
    \end{tabular}
    \label{tab:intuition_keywords}
\end{table}

\paragraph{Computation of EMI score.} In our methodology, we start by training 300-dimensional word embeddings using the Word2Vec \cite{NIPS2013_9aa42b31} algorithm on the corpus of Congressional speeches. We use the Gensim library \cite{rehurek_lrec}. Word2Vec is an algorithm that generates dense vector representations of words, known as word embeddings. The rationale behind using Word2Vec lies in its ability to capture semantic relationships among words by representing them in a continuous vector space. This algorithm learns to predict the context of a word based on its surrounding words or vice versa. The resulting word embeddings encode semantic similarities, making them valuable for computational analysis of language. 

Following this, we compute a representation for the concepts of interest by averaging the word embeddings for the relevant keywords in the respective dictionaries for evidence-based and intuition-based language. For a given text, we compute its representation by taking the average of the word embeddings for its content words. 
This representation allows for a graded measure of relatedness to each construct as we can calculate the cosine similarity between each construct representation and the representation of a given target text that is computed in the same manner. 

To generate the representations and compute cosine similarities, we use the sentence-transformers library \cite{reimers-2019-sentence-bert}, leveraging our trained Word2Vec model. This approach offers efficiency and effectiveness in capturing the semantic content of textual data. This setup allows us to obtain textual embeddings with minimal computational resources and ensures the scalability of our analysis. 

To address variations in the length of speeches, we perform length adjustments
for the cosine similarities. This involves binning the similarities by length and subtracting the mean similarity within each bin from the cosine similarity of each instance. Subsequently, we apply a Z-transform to the cosine similarities to derive the evidence and intuition scores. Finally, we obtain the Evidence-Minus-Intuition (EMI) score by subtracting the intuition score from the evidence score. A positive EMI score indicates a higher prevalence of evidence-based language whereas
a negative score suggests a reliance on intuition-based language. Tables \ref{tab:pos_emi} and \ref{tab:neg_emi} contain illustrative examples of speeches with postive and negative EMI score, repectively. For further analysis, we take the mean of the EMI score per two-year period, corresponding to the typical duration of Congressional sessions.

\paragraph{Validation of EMI score over time.}
 We split the EMI score into 4 bins per decade. We sample 5 (4 for the most recent decade) (quasi)sentences from each bin per party (Democrats vs. Republicans) and decade, resulting in a sample of size 592. We ask participants on Prolific to rate to what extent a given text is evidence-based and intuition-based (or evidence-free) on two Likert scales ranging from 1 to 7. Each text has at least 5 ratings. We collected a total of 4,563 human ratings from 156 participants. The average number of ratings provided by each participant is 29 (with a minimum of 11 and maximum of 30). As the average of the ratings for each scale are negatively correlated at the document level ($-0.85$, $p<0.001$), we derive human judgement by assigning a label of evidence-based if the average evidence-based rating is greater than the average intuition-based rating, otherwise we classify the item as intuition-based. Annotators have relatively high levels of agreement. Sampling five annotations at random for each text, the Intra Class Correlation for the mean of the difference between the evidence-based and intuition-based scales is $0.714$ (95\% CI = $[0.675,0.749]$).
 
 We calculate the Area Under the Curve (AUC) of the receiver operating characteristics curve as the evaluation metric following previous work that used classification metrics \cite{gennaro2022emotion, lasser2023alternative}. We compute the AUC score per decade and for all samples. The AUC is a measure of the reliability of our computed EMI score, with a score of 1 indicating perfect accuracy and 0.5 representing performance equivalent to random chance. Our method achieves an overall AUC of 0.79 across decades, ranging from 0.60 to 0.94 (see Table \ref{tab:auc_perdecade}). Compared to the random baseline AUC of 0.5, our method demonstrates acceptable to excellent discrimination levels \cite{hosmer2013applied}.
 
\begin{table}[!htbp]
    \centering
    \caption{AUC per decade and overall AUC computed on the full sample without temporal split }
    \begin{tabular}{|ccc|} 
         \hline
         Decade starting&AUC &Number of speeches\\ \hline
         1879&0.81& 40\\ 
         1889&0.60& 40\\ 
         1899&0.61& 40\\ 
         1909&0.82& 40\\ 
         1919&0.61& 40\\ 
         1929&0.83& 40\\ 
         1939&0.82& 40\\ 
         1949&0.82& 40\\ 
         1959&0.93& 40\\ 
         1969&0.93& 40\\ 
         1979&0.90&40\\
 1989&0.94&40\\
 1999&0.93&40\\
 2009&0.83&40\\
 2019&0.74&32\\ \hline
         Overall&0.79&592\\ \hline
    \end{tabular}
    \label{tab:auc_perdecade}
\end{table}

\begin{table}[!htbp]
    \centering
    \caption{Examples of speeches with positive EMI score}
    \renewcommand{\arraystretch}{1.5}
    \begin{tabular}{|p{\dimexpr\textwidth-2\tabcolsep-2\arrayrulewidth}|}
           \hline
           \multicolumn{1}{|c|}{Sample speeches} \\ \hline
         ``Lives have been directly affected- by the tragedy of suicide were also called upon to provide insight into what might be done in the family and in the community to prevent the further senseless waste of young lives. By all accounts, the conference was a tremendous success. In fact, many participants returned to their communities and, with the knowledge obtained from the conference, established suicide prevention programs. To assist other communities, the Youth Suicide National Center, in conjunction with the Office of Human Services, Administration for Children, Youth, and Families, of the Department of Health and Human Services, has compiled for dissemination the findings and recommendations of the conference. I note that the findings and recommendations will be published within 1 year of the conference, thereby recognizing the urgency associated with the problem. In sum, the administration has been involved in an effort to address the tragedy of youth suicide'' \\
        ``Yes. The Senator can get them in detail I am sure from the report of the Federal Trade investigation. Before I conclude I shall give some figures as to some of the holding companies and subsidiaries, and some figures applying to all of them showing the fictitious capital stocks and bonds which have been floated and sold to an innocent public, for which absolutely no real value existed.'' \\
        ``If the distinguished and honorable Senator has not read the report of the committee, which was prepared for the information of Senators, he must not charge the committee with any dereliction of duty in not supplying him with information.'' \\
         ``Is it not true that the only basis of valuation that can be established for a fixed rate of return is through the property investment account until the actual value is ascertained by the physical valuation of the railroads by the Interstate Commerce Commission?'' \\
       ``May I ask the Senator from Iowa to repeat the form in which the comparison has been tabulated? My attention was diverted to another quarter at the moment. In what shape will the comparative table be?'' \\ \hline         
    \end{tabular}
    \label{tab:pos_emi}
\end{table}

\begin{table}[!htbp]
    \centering
     \caption{Examples of speeches with negative EMI score}
    \renewcommand{\arraystretch}{1.5}
    \begin{tabular}{|p{\dimexpr\textwidth-2\tabcolsep-2\arrayrulewidth}|}
           \hline
           \multicolumn{1}{|c|}{Sample speeches}  \\ \hline
          ``I can give the Senator an illustration. I had some ancestors who were very smart people. but fought for the Stuarts in Great Britain against Puritanism and the Commonwealth and the Parliament. They were wise men individually. but historically they were asses. Does the Senator understand the illustration? Their successors partially in my person have confessed that they were asses.'' \\
       ``I desire to say to the gentleman from Wisconsin. who never "shakes his head wisely." that he has no business to shake it unwisely at me.'' \\
     ``Oh. yes. your howl about the farmers of the country and the destruction of the price of wheat is nothing but the wail of the old standpatter. who sees the mountain of protection giving way under ceaseless and constant hammering on the part of the people.'' \\
    ``Mr. Speaker. I join with my colleagues in expressing my sorrow at the passing of our former colleague. the Honorable Charles A. Halleck of Indiana. It was my privilege to serve with Charlie for quite a long time. The leader on the Republican side. he was a tireless worker. candid. honest. and able. He contributed greatly to his country and to the Congress itself. May I say that his work here will be long remembered and his contributions to his district. State. and Nation will be lasting.'' \\
    ``Mr. Speaker, in the face of this impasse and in the spirit of the season. I believe we should forget this Democrat versus Republican stuff, legislators versus Executive Branch, liberals versus conservatives, and unite under the common bond of being Americans. We are reminded of a similar impasse in our history at the constitutional convention when the sage elder statesman, Ben Franklin, stood with these words: "In the beginning of our war with Britain. we prayed daily for guidance. Our prayers were heard and answered. Have we now forgotten this powerful ally? The longer I live, this I know to be true, God governs the affairs of men, for if a sparrow cannot fall without His notice, is it probable that a nation can rise without His aid? The psalmist tells us in chapter 118. verse 8. "Put your trust in God. not confidence in men." We have these same words above the Speakers chair and right over the American flag. I believe that we, as a Congress, should come together as Democrats and Republicans and leaders to do what is best for the American country, put God and country first, not partisan, politics.'' \\
    \hline         
    \end{tabular}   
    \label{tab:neg_emi}
\end{table}

\newpage
\paragraph{Example speeches in periods with low overall EMI score.}
Table \ref{tab:sample_periods} shows examples of speeches with low EMI score (in the bottom 1\%) in periods with overall low EMI score in Figure~\ref{fig:trend_panel}A. Consistent with previous research~\cite{pennebaker2003psychological, wallace1993political, pennebaker2002language} that highlighted
changes in language of individuals and political leaders during crises, these examples suggest a tendency for discussions about the crisis of the time to rely more on intuition-based language rather than evidence-based language.

\begin{table}[hbtp]
    \centering
    \caption{Example of speeches with low EMI score in specific periods. There is a tendency to focus on prevailing crisis, including the war in the Philippines in the 1890s, the impact of the Great Depression in the 1930s, and challenges associated with raising the debt ceiling in the 2020s.}
    \renewcommand{\arraystretch}{1.5}
    \resizebox{\textwidth}{!}{
    \begin{tabular}{|c| p{\dimexpr\textwidth-9\tabcolsep-9\arrayrulewidth}|}
         \hline
         Period & \multicolumn{1}{|c|}{Sample speeches}  \\ \hline
         1899 & ``Philippine policy whatever that may be right or wrong. is the veriest rot. an insult to intelligence. a shame upon manhood. a tale told by an idiot. a betrayal of the principle of selfgovernment. I am willing to go as far as anyone in patriotism. I will support the country in any emergency. but President McKinley is not the country. The time has not yet comeI pray Almighty God that it may never arisewhen the American people will accept the arrogant dictum of Louis XIV. when repeated by an American President. "I am the State!" If President McKinley is at all worthy of his high position. he must entertain a supreme contempt for those political invertebrates. particularly for those. claiming to be Democrats. who. in order to catch the crumbs falling from their masters table [applause]. go about saying. "The President is wrong in his Philippine policy. but we must support'' \\
         1933 & ``expression of such noble sentimentsI find myself utterly unable to reconcile his refusal to relieve the misery. the fears. the want. the privation. the bitter humiliation of millions of our finest men and women and children of this land. To me it is an utterly incomprehensible reversal of everything Franklin Delano Roosevelt has voiced in the past. everything he has claimed to stand for. everything he has typified to a Nation whose citizens today still stand on the perilous edge of a yawning precipice. into the darkness and social chaos of which a thrust. the like of that which the President has made into the hearts and the hopes of millions of Americans. might easily plunge us. I want to say to you. Mr. Chairman. that in the person of Franklin Delano Roosevelt. into his keeping. because of the high spiritual ideals he has voiced. the high spiritual promises he'' \\
         2021 & ``aside differences and move it forward. But now, unfortunately, some of our Republican colleagues even though they were eager to have Democrats support them when President Trump was President now some of our Republican colleagues are reportedly contemplating a reckless idea, spearheaded by the Republican Senator from Wisconsin, to oppose any effort to raise the debt ceiling whatsoever. And, unfortunately and sadly, the Republican leader seems to be going along. Let me be clear: taking the debt hostage and playing games with the full faith and credit of the United States is reckless, irresponsible, and will harm every single American. It is a complete nonstarter. This is not just another political debate. It is about honoring our unbroken commitment to pay our debts and avoid another financial crisis at a crucial moment for our country. Now it is important to remember that this is not about green-lighting future spending'' \\ \hline
    \end{tabular}
    }
     \label{tab:sample_periods}
\end{table}

\begin{figure}[!hbtp]
    \centering
    \includegraphics[width=\textwidth, height=0.9\textheight, keepaspectratio]{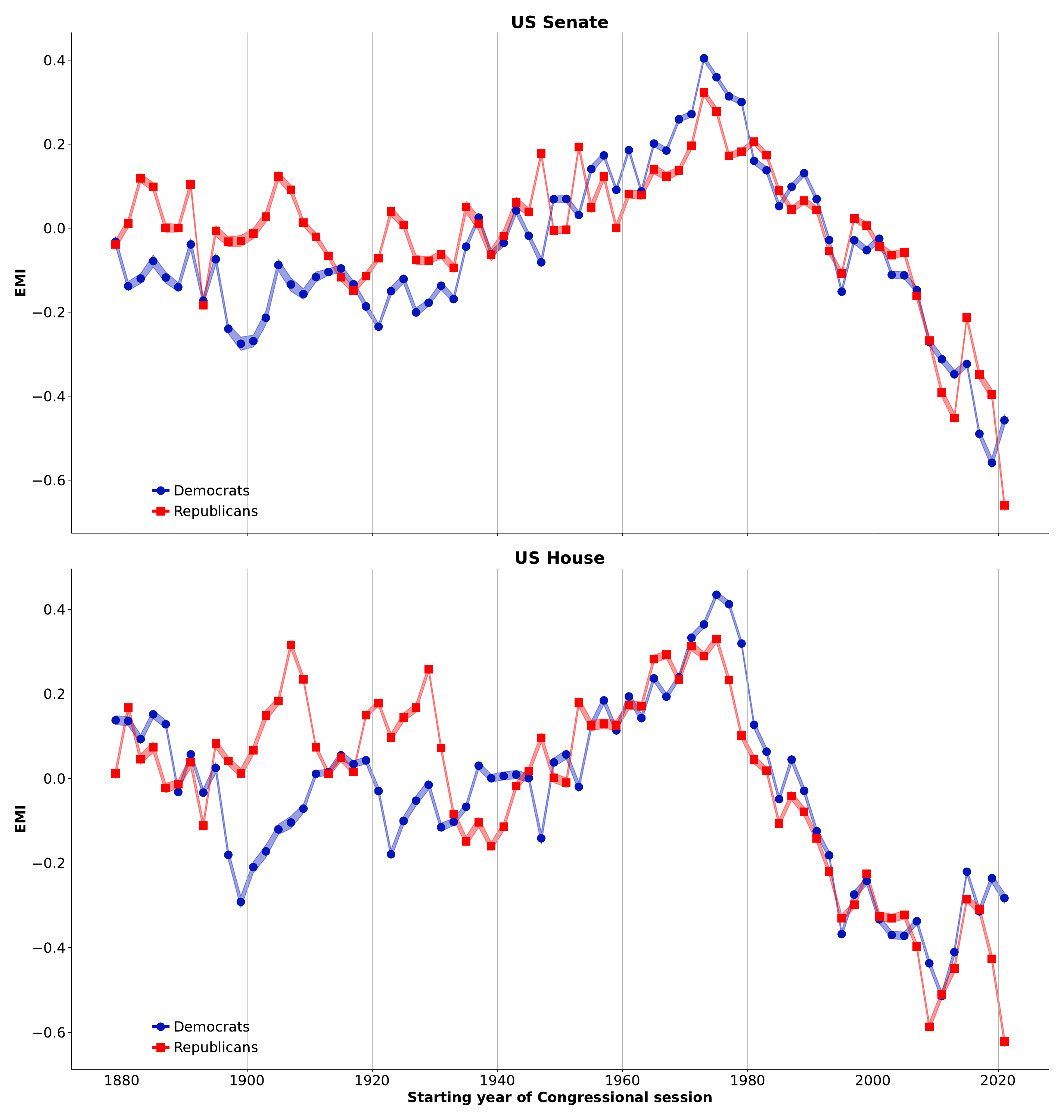}
    \caption{Time series of EMI by party in the U.S. Senate and House}
    \label{fig:emi_party_chamber}
\end{figure}

\paragraph{EMI in the House and Senate by party.}
Figure \ref{fig:emi_party_chamber} shows the trend of EMI by party in both chambers of the U.S. Congress over time. The trends follow a similar pattern to the one observed for the overall EMI score in the main text in panels A and B in Figure~\ref{fig:trend_panel}.

\newpage
\subsection*{Statistical analysis of the trends in EMI, polarization, and inequality}

We fit time series as linear regression models that include lagged dependent variables to consider autocorrelation. For each time series, we fit autoregressive (AR) models with increasing lags up to a point in which the quality of models does not improve with additional lags. In all cases we report, inclusion
of one lag generated the best univariate AR model. We next extend these models with other variables including EMI and other covariates. We measure variance inflation factors (VIF) of the independent variables of the models and include interaction terms when any of the covariates, excluding the lagged dependent variable, has a VIF above 10. 
After fitting a model specified in this way, we measure standard errors and p-values with a Heteroskedasticity and Auto-Correlation (HAC) adjusted estimator. 
We assess the stationarity of residuals with Augmented Dickey-Fuller (ADF) tests and 
Kwiatkowski-Phillips-Schmidt-Shin (KPSS) tests,
and the normality of residual distributions with Jarque-Bera (JB) tests. 
Models generally passed these regression diagnostics, being able to reject the null
hypothesis of the ADF test at a 0.05 level and failing to reject the null of the
KPSS and JB tests at a 0.1 level. 
We report here any relevant cases where those diagnostics are different.

In our primary analysis of inequality, we consider the fraction of income of the top 1\% from 1944, which is the year when tax declaration exemption rules qualitatively changed and led to more reliable inequality metrics \cite{piketty2003income}. We assessed the robustness of our results with alternative specifications, namely using the Gini index (from \url{https://www.census.gov/data/tables/time-series/demo/income-poverty/historical-income-inequality.html}) in the same period and using the full record of the share of income of the top 1\% from 1912. 

\begin{figure}[!htbp]
    \centering
    \includegraphics[width=0.95\textwidth, height=0.9\textheight]{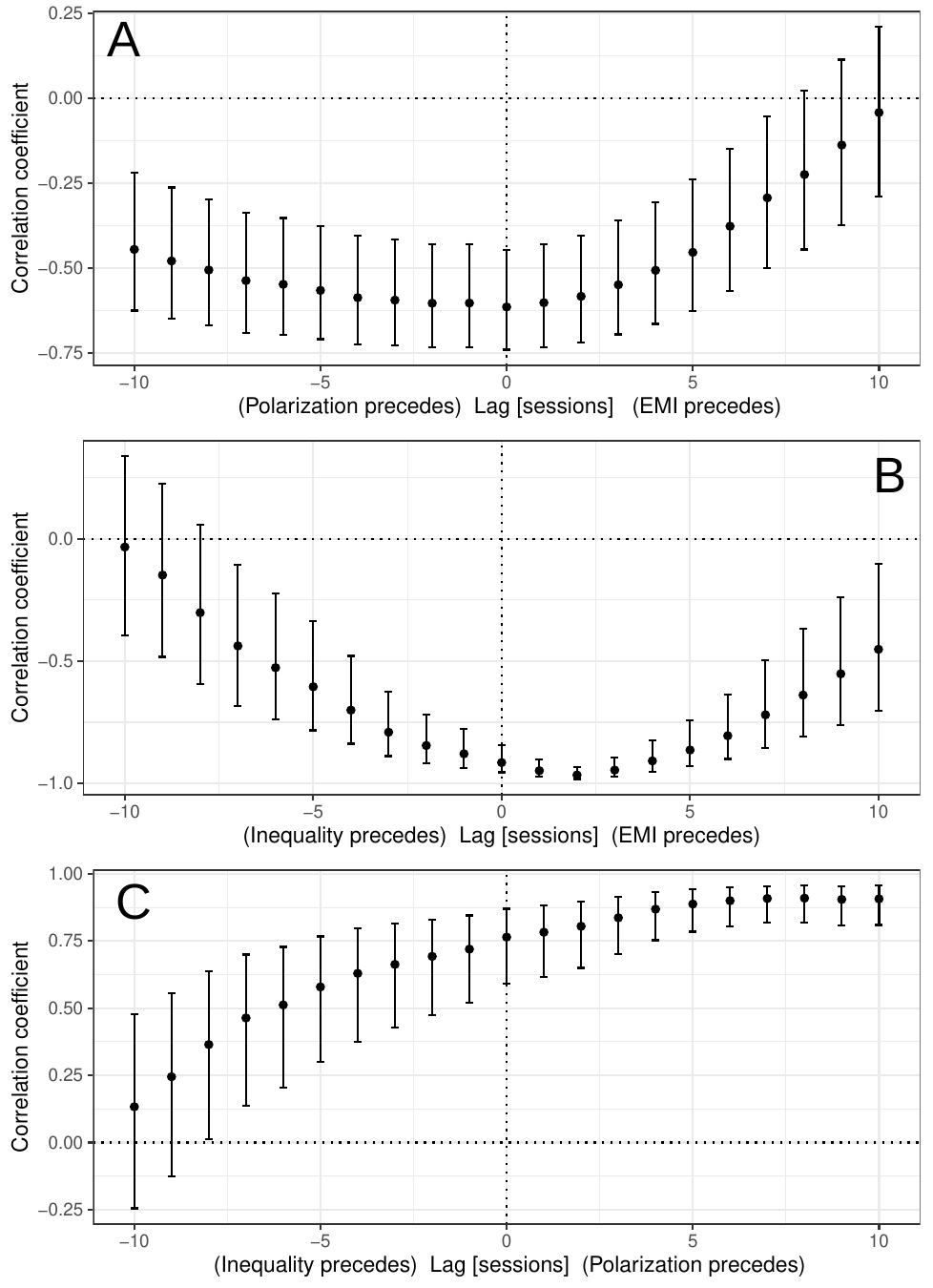}
    \caption{Lagged correlation analysis between EMI and Polarization (A), EMI and Inequality (B) and Inequality and Polarization (C). 95\% confidence intervals are shown.}
    \label{fig:crosscorr}
\end{figure}

We add one more specification to robustly test how the role of polarization influences our results about EMI and inequality. A lagged correlation analysis between inequality and polarization indicates that the correlation between these two is strongest when considering a lag of 8 legislative sessions (see Figure \ref{fig:crosscorr}C). To consider this longer lag, we fitted an additional regression model of inequality with EMI and the previous value of inequality, but with the value of polarization 8 sessions prior. Results of this fit are reported in table \ref{tab:EMI-Ineq-Ext}.

\begin{figure}[!htbp]
    \centering
    \includegraphics[width=0.95\textwidth]{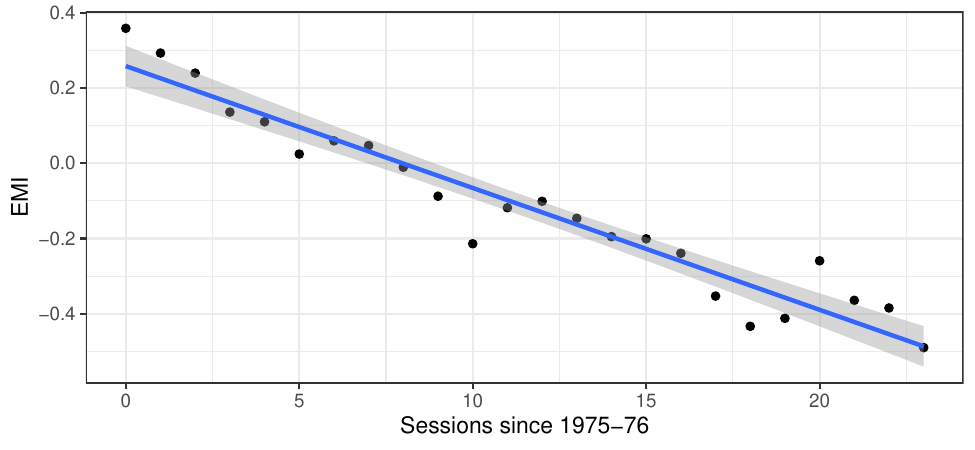}
    \caption{EMI values after the peak in 1975-76. The line shows a linear fit with a shaded area showing a standard error around the predicted value of EMI.}
    \label{fig:EMItrend}
\end{figure}

\paragraph{EMI negative trend.}
The session with the highest EMI score is 1975-76, with a score of 0.358, closely followed by the previous session with an EMI score of 0.355 but substantially higher than the mean session, which has a slightly negative EMI of -0.017. The peak EMI is more than two standard deviations above the mean of the historical distribution (sd=0.174). From that peak, a downwards trend is noticeable and is confirmed by a linear regression model of the form:
\begin{align*}
    EMI(t) = a + b \times t
\end{align*}
The fit has an intercept $a= 0.258$ and a slope $b=-0.032$, both with $p<0.001$. The model has $R^2=0.927$ and the fit can be seen in Figure \ref{fig:EMItrend}.

\paragraph{EMI and polarization.}
To measure partisan polarization in Congress, we use the first dimension of the DW-NOMINATE (Dynamic Weighted NOMINAl Three-step Estimation) score \cite{poole2011ideology}, which measures the ideological position of members of Congress derived from their roll-call votes. 
The difference in aggregate score for the two major parties reflects the extent to which they differ ideologically. A higher difference in the first dimension indicates a greater ideological distance or polarization between the parties. A lower difference suggests a closer alignment in their ideological positions. We use the DW-NOMINATE data from VOTEVIEW \cite{lewis2021voteview} which offers a comprehensive and widely used resource for studying the ideological landscape and partisan dynamics within the U.S. Congress.

\begin{figure}[!htbp]
    \centering
    \includegraphics[width=\textwidth]{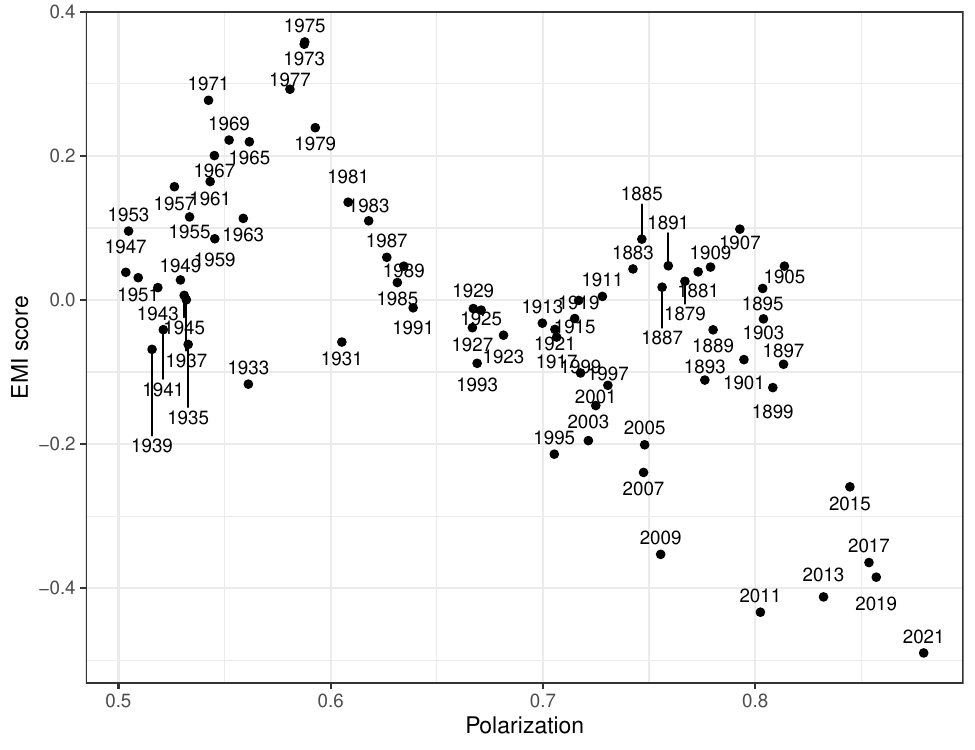}
    \caption{Scatter plot of polarization and EMI in the same year. The cross-correlation between these two variables is -0.615 (95\% $CI = [-0.741,-0.447]$, $p<0.01$)}
    \label{fig:pol_emi_scatter}
\end{figure}

To understand the relationship between EMI and polarization (scatter plot in Figure \ref{fig:pol_emi_scatter}), we fitted lagged regression models of the form:
\begin{align*}
    EMI(t) = a + b \times EMI(t-1) + c \times Pol (t-1)
\end{align*}
\begin{align*}
    Pol(t) = a + b \times Pol(t-1) + c \times EMI (t-1)
\end{align*}
and compared them to autoregressive models ignoring the other variable. Results of the fits show that polarization does not have a significant coefficient in the EMI model and that the polarization model has a significant negative coefficient for EMI, but of small magnitude when compared to the autoregressive coefficient of EMI. 
A KPSS test of residuals in this model rejects the null hypothesis ($p=0.036$) but an ADF test also rejects the null ($p=0.02$). While residuals deviate a bit from being stationary, we use HAC covariance matrix estimation and residuals do not significantly deviate from normality, as a Jarque-Bera test is not significant ($p=0.645$).

\begin{table}[!htbp] \centering 
  \caption{Models of temporal effects between Polarization and EMI.} 
  \label{tab:EMI-Pol} 
  \resizebox{\textwidth}{!}{
\begin{tabular}{@{\extracolsep{5pt}}lcccc} 
\\[-1.8ex]\hline 
\hline \\[-1.8ex] 
 & \multicolumn{4}{c}{\textit{Dependent variable:}} \\ 
\cline{2-5} 
\\[-1.8ex] & \multicolumn{2}{c}{EMI} & \multicolumn{2}{c}{Pol} \\ 
\hline \\[-1.8ex] 
 EMI(t-1) & \textbf{0.98}$^{**}$ & \textbf{0.92}$^{**}$ &  &  \\ 
  & (0.06) & (0.07) &  &  \\ 
 Pol &  & \textbf{$-$0.15}$^{*}$ &  &  \\ 
  &  & (0.07) &  &  \\ 
 Pol(t-1) &  &  & \textbf{1.00}$^{**}$ & \textbf{0.97}$^{**}$ \\ 
  &  &  & (0.02) & (0.03) \\ 
 EMI &  &  &  & $-$0.03 \\ 
  &  &  &  & (0.02) \\ 
 Intercept & $-$0.01 & 0.09$^{*}$ & $-$0.0002 & 0.02 \\ 
  & (0.01) & (0.05) & (0.02) & (0.02) \\ 
\hline \\[-1.8ex] 
Observations & 71 & 71 & 71 & 71 \\ 
R$^{2}$ & 0.86 & 0.87 & 0.96 & 0.97 \\ 
Adjusted R$^{2}$ & 0.86 & 0.87 & 0.96 & 0.97 \\ 
Residual Std. Error & 0.06 & 0.06 & 0.02 & 0.02 \\ 
F Statistic & 438.71$^{**}$ & 227.64$^{**}$ & 1,887.01$^{**}$ & 967.03$^{**}$ \\ 
\hline 
\hline \\[-1.8ex] 
\textit{Note: Numbers in parentheses represent standard errors.}   & \multicolumn{4}{r}{$^{\dag}$p$<$0.1; $^{*}$p$<$0.05; $^{**}$p$<$0.01} \\ 
\end{tabular} 
}
\end{table} 

\newpage
\paragraph{EMI and inequality.}
To measure income inequality, we use the share of pre-tax income of the top 1\% of the population \cite{alvaredo2016distributional}. The data is from the world inequality database (\url{https://wid.world/}).
A lagged correlation analysis shows that the strongest correlation between EMI and inequality has a lag 2, where EMI precedes inequality (see panel B of Figure \ref{fig:crosscorr}). Inequality is also known to be correlated with polarization \cite{mccarty2016polarized}, which we also observe in our lagged correlation analysis in panel C of Figure \ref{fig:crosscorr}. For that reason, we study the role of EMI in inequality while considering polarization, as EMI and polarization are negatively cross-correlated. 
The Variance Inflation Factor of a specification including lagged measures of inequality, EMI, and Polarization is 9.67, indicating that we need to include an interaction term between EMI and Polarization. Thus, our model has the form:

\begin{align*}
            Ineq(t) = a + b \times Ineq(t-1) + c \times EMI(t-1) + d \times Pol(t-1) \\ 
        + e \times EMI(t-1) \times Pol(t-1)        
\end{align*}

\begin{figure}
    \centering
    \includegraphics[width=\textwidth]{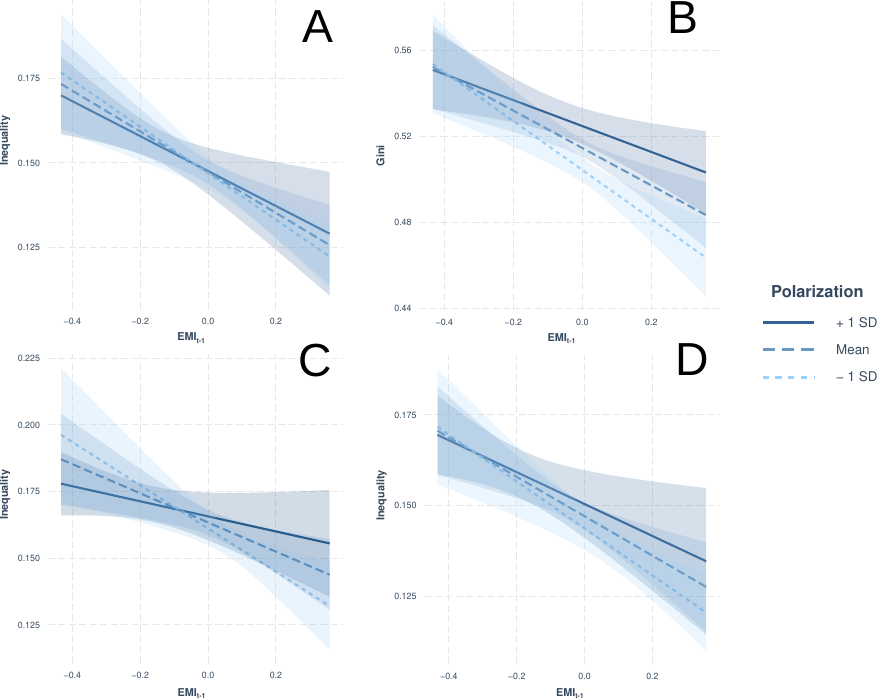}
    \caption{Interaction plots of models of Inequality as a function of lagged values of inequality, polarization, and EMI. The figure focuses on the relationship between inequality and EMI in the previous session for three levels of polarization in the previous session: the mean and one standard deviation above and below. The shaded area around regression lines are standard errors. Panel A shows our first specification, panel B shows the result for a model with Gini index as a measure of inequality, panel C shows the result including less reliable inequality data since 1912, and panel D shows a version of our original model but including polarization values with a lag of 8.}
    \label{fig:interactions}
\end{figure}

We compare this model to a simple autoregressive model including lagged values of inequality and polarization. The result is shown in Table \ref{tab:EMI-Ineq}. The lagged value of EMI has a negative and significant coefficient on inequality and the interaction with Polarization is not significant. Knowing the EMI in one session improves the prediction of inequality in the two-year period that follows. The interaction between EMI and polarization, while positive, does not lead to an important mediation in the role of EMI, as shown in Figure \ref{fig:interactions}A.
Residuals in this model are stationary (ADF $p=0.022$, KSPP $p>0.1$) and do not deviate from normality (JB $p=0.822$).

The results of our analysis of inequality remain qualitatively similar with different specifications for the decisions we took in our analysis above. Table \ref{tab:EMI-Ineq-Ext} shows the results, where the first model uses the Gini index as a measure of inequality.  
In this specification, the VIF of predictors is 13.17, indicating the need for inclusion of an interaction term between polarization and EMI. The coefficient of EMI is negative and significant, but it has a significant positive interaction with polarization. Figure \ref{fig:interactions}B shows the shape of this interaction, revealing that high levels of polarization do not reverse the effect of EMI. Residuals in this model are stationary (ADF $p<0.01$, KPSS $p>0.1$) and do not deviate from normality (JB $p=0.343$).
The second model uses the share of income of the top 1\% of the population but includes less reliable data since 1912. In this model, the VIF is 3.175, but we keep the interaction term between polarization and EMI for comparability to other models. The result is similar as for the case using the Gini index: the coefficient for EMI is negative and significant but the interaction with polarization is positive and significant. Figure \ref{fig:interactions}C shows the shape of this interaction, where high levels of polarization do not reverse the slope of Inequality with EMI. Residuals in this model are stationary (ADF $p < 0.01$, KPSS $p= 0.09$) but deviate from normality (JB $p<0.01$). For that reason, we performed a bootstrapping test on the coefficient of EMI with 10,000 samples, which indicates that the negative coefficient for EMI is robust to non-normal residuals (95\% CI=$[-0.34,-0.13]$).
The model with a longer lag for polarization also has a high VIF of 10.02, motivating the inclusion of the interaction between EMI and polarization. Residuals are stationary (ADF $p<0.01$, KPSS $p>0.1$) and do not deviate from normality (JB $p=0.716$). In this model, the coefficient of EMI is also negative and significant and the interaction between polarization and EMI is only significant at the 0.1 level.
Figure~\ref{fig:interactions}D shows the shape of this interaction, revealing the same pattern in which even for high polarization, the slope of EMI is negative.

\begin{table}[!htbp] \centering 
  \caption{Regression results of models of inequality as a function of lagged values of inequality, polarization, EMI, and the interaction between EMI and polarization.} 
  \label{tab:EMI-Ineq} 
\begin{tabular}{@{\extracolsep{5pt}}lcc} 
\\[-1.8ex]\hline 
\hline \\[-1.8ex] 
 & \multicolumn{2}{c}{\textit{Dependent variable:}} \\ 
\cline{2-3} 
\\[-1.8ex] & \multicolumn{2}{c}{Ineq} \\ 
\hline \\[-1.8ex] 
 Ineq(t-1) & \textbf{0.87}$^{**}$ & \textbf{0.57}$^{**}$ \\ 
  & (0.08) & (0.08) \\ 
 Pol(t-1) & \textbf{0.04}$^{*}$ & 0.003 \\ 
  & (0.02) & (0.02) \\ 
 EMI(t-1) &  & \textbf{$-$0.11}$^{**}$ \\ 
  &  & (0.03) \\ 
 Pol(t-1)*EMI(t-1) &  & 0.08 \\ 
  &  & (0.05) \\ 
 Intercept & $-$0.01 & 0.06$^{**}$ \\ 
  & (0.01) & (0.02) \\ 
\hline \\[-1.8ex] 
Observations & 38 & 38 \\ 
R$^{2}$ & 0.92 & 0.95 \\ 
Adjusted R$^{2}$ & 0.92 & 0.95 \\ 
Residual Std. Error & 0.01 & 0.01 \\ 
F Statistic & 212.40$^{**}$ & 160.28$^{**}$ \\ 
\hline 
\hline \\[-1.8ex] 
\textit{Note: Numbers in parentheses represent standard errors.}  & \multicolumn{2}{r}{$^{\dag}$p$<$0.1; $^{*}$p$<$0.05; $^{**}$p$<$0.01} \\ 
\end{tabular} 
\end{table}

\begin{table}[!htbp] \centering 
  \caption{Regression results for alternative specifications of our inequality analysis. Model 1 uses the Gini index as a measure of inequality, Model 2 uses all data on the share of income of the 1\% since 1912, and Model 3 uses a longer lag of 8 legislative sessions for polarization.} 
  \label{tab:EMI-Ineq-Ext} 
\resizebox{\textwidth}{!}{
\begin{tabular}{@{\extracolsep{5pt}}lccc} 
\\[-1.8ex]\hline 
\hline \\[-1.8ex] 
 & \multicolumn{3}{c}{\textit{Dependent variable:}} \\ 
\cline{2-4} 
\\[-1.8ex] & Ineq (Gini) & Ineq (since 1912) & Ineq \\ 
\\[-1.8ex] & (1) & (2) & (3)\\ 
\hline \\[-1.8ex] 

 Ineq (t-1)& \textbf{0.46}$^{**}$ & \textbf{0.70}$^{**}$ & \textbf{0.52}$^{**}$ \\  & (0.10) & (0.10) & (0.09) \\ 

 EMI(t-1) & \textbf{$-$0.24}$^{**}$ & \textbf{$-$0.22}$^{**}$ & \textbf{$-$0.14}$^{**}$ \\  & (0.05) & (0.05) & (0.04) \\ 
 
 Pol(t-1) & \textbf{0.09}$^{**}$ & 0.02 &  \\   & (0.03) & (0.04) &  \\ 
 
 Pol(t-1)*EMI(t-1) & \textbf{0.25}$^{**}$ & \textbf{0.26}$^{**}$ &  \\   & (0.06) & (0.07) &  \\ 
  
 EMI(t-1)*Pol(t-8) &  &  & 0.14$^{\dag}$ \\ &  &  & (0.08) \\ 
 
 Pol(t-8) &  &  & 0.04 \\   &  &  & (0.04) \\ 
 
 Intercept & 0.22$^{**}$ & 0.04 & 0.04$^{*}$ \\   & (0.04) & (0.03) & (0.02) \\ 

\hline \\[-1.8ex] 
Observations & 38 & 54 & 31 \\ 
R$^{2}$ & 0.97 & 0.87 & 0.97 \\ 
Adjusted R$^{2}$ & 0.97 & 0.86 & 0.96 \\ 
Residual Std. Error & 0.01 & 0.01 & 0.01 \\ 
F Statistic & 315.53$^{**}$ & 81.45$^{**}$ & 195.75$^{**}$ \\ 
\hline 
\hline \\[-1.8ex] 
\textit{Note: Numbers in parentheses represent standard errors.}  & \multicolumn{3}{r}{$^{\dag}$p$<$0.1; $^{*}$p$<$0.05; $^{**}$p$<$0.01} \\ 
\end{tabular} 
}
\end{table} 

\newpage
\subsubsection*{Statistical analysis of the relationship between EMI and measures of Congressional productivity}

Following the specification of \cite{grant2008legislative}, we fit a base model of three congressional productivity indices (MLI, LPI, and log-transformed number of laws) as a function of the lagged dependent variable, polarization, policy mood, and two indicator variables for whether the same party controls both the Presidency and the majority in Congress and for a change in this variable. We extend this model by adding the EMI score of the same session in which productivity is measured. Thus, our model is for each variable Y (MLI, LPI, log laws):
\begin{align*}
 Y(t) = a + b \times Y(t-1) + c \times Pol(t) + d \times Mood(t)
 + e \times PartyControl(t) \\
 + f \times PartyControlDif(t) + g \times EMI(t) + h \times EMI(t) \times Pol(t)   
\end{align*}
Note that in this model specification, we use the EMI in the same session as the congressional productivity metric, as we are trying to identify a correlation between variables that is robust to the known effect of other indicators. Table \ref{tab:Prod} shows regression results.
Across the three models, explanatory variables reached VIF values up to 12.98, so we included an interaction term between polarization and EMI. Tests of stationarity of residuals had lower significance due to the smaller sample sizes (ADF p=0.26 for MLI, p=0.07 for LPI, and p=0.03 for number of laws), but KPSS tests were not significant in all three cases (p$>$0.1) and JB tests were not significant either (p$>$0.5). These small deviations from stationarity of residuals are corrected with the HAC covariance estimator.

While our analysis of productivity includes the important variable of \textit{Mood}, data on public policy mood is only available since the 1950s, as it was collected via surveys. To analyze further the role of EMI in productivity, we adopt the approach of \cite{grant2008legislative}, using the logarithm of the number of patents (from~\url{https://www.uspto.gov/web/offices/ac/ido/oeip/taf/h_counts.htm}) approved during each session as an approximation of public mood regarding regulation. While this is an imperfect approximation, it allows us to study a much longer period, dating back to the 19th century. Thus, for each dependent variable, we now have models of the form:
\begin{align*}
 Y(t) = a + b \times Y(t-1) + c \times Pol(t) + d \times npatents(t) + e \times PartyControl(t) \\
 + f \times PartyControlDif(t) + g \times EMI(t) + h \times EMI(t) \times Pol(t)   
\end{align*}
where $npatents(t)$ represents the logarithm of the number of patents approved during the congressional session $t$. Covariates in this model have VIF up to 7.67 (LPI), and therefore we include an interaction term between EMI and Pol in each model. Results are shown in Table \ref{tab:Prod2}. Residuals were approximately stationary, with significant ADF tests for MLI (p=0.014) and number of laws (p=0.01), and significant at the 0.1 level for LPI (p=0.09). KPSS tests were not significant for all three models (p$>$0.1) and JB tests were not significant for LPI (p=0.33) and number of laws (0.52). For MLI, a JB test was significant (p$<$0.01), indicating non-normal residuals. For that reason, we performed a bootstrap test with 10,000 samples, which gave a 95\% Confidence Interval for the coefficient of EMI of $[0.026,0.214]$, indicating that the significant  coefficient of the MLI model is robust to deviations from normality in the residuals. The coefficients of interaction terms between EMI and polarization are not significant and the coefficient for EMI is only significant for MLI, while it is not for LPI nor the number of laws.

\begin{table}[htbp] \centering 
  \caption{Models of Congressional productivity as a function of EMI and relevant covariates. Fits start in 1951 to include policy mood data and end in 2004 for MLI and LPI and in 2022 for the logarithm of the number of laws passed in a session.} 
  \label{tab:Prod} 
\resizebox{\textwidth}{!}{
\begin{tabular}{@{\extracolsep{5pt}}lcccccc} 
\\[-1.8ex]\hline 
\hline \\[-1.8ex] 
 & \multicolumn{6}{c}{\textit{Dependent variable:}} \\ 
\cline{2-7} 
\\[-1.8ex] & \multicolumn{2}{c}{MLI} & \multicolumn{2}{c}{LPI} & \multicolumn{2}{c}{nlaw} \\ 
\\[-1.8ex] & (1) & (2) & (3) & (4) & (5) & (6)\\ 
\hline \\[-1.8ex] 
 MLI(t-1) & \textbf{0.95}$^{**}$ & \textbf{0.77}$^{**}$ &  &  &  &  \\ 
  & (0.06) & (0.07) &  &  &  &  \\ 
 EMI(t) &  & \textbf{0.67}$^{*}$ &  & \textbf{0.83}$^{**}$ &  & 0.27$^{\dag}$ \\ 
  &  & (0.25) &  & (0.21) &  & (0.16) \\ 
 LPI(t-1) &  &  & \textbf{0.75}$^{**}$ & \textbf{0.44}$^{**}$ &  &  \\ 
  &  &  & (0.09) & (0.08) &  &  \\ 
 nlaw(t-1) &  &  &  &  & $-$0.24 & $-$0.21 \\ 
  &  &  &  &  & (0.15) & (0.16) \\ 
 Pol(t) & \textbf{$-$0.31}$^{**}$ & 0.21 & \textbf{$-$0.45}$^{*}$ & $-$0.14 & \textbf{$-$1.16}$^{**}$ & \textbf{$-$0.89}$^{**}$ \\ 
  & (0.08) & (0.24) & (0.16) & (0.16) & (0.14) & (0.26) \\ 
 PartyControl(t) & $-$0.16 & $-$0.08 & $-$0.14 & $-$0.07 & $-$0.10 & $-$0.06 \\ 
  & (0.17) & (0.19) & (0.14) & (0.14) & (0.20) & (0.20) \\ 
 PartyControlDif(t) & 0.19 & 0.27 & 0.23 & \textbf{0.29}$^{*}$ & 0.01 & 0.05 \\ 
  & (0.20) & (0.19) & (0.18) & (0.14) & (0.16) & (0.15) \\ 
 Mood(t) & \textbf{0.17$^{*}$} & \textbf{0.24$^{**}$} & \textbf{0.17$^{*}$} & \textbf{0.30$^{**}$} & \textbf{0.37$^{**}$} & \textbf{0.39$^{**}$} \\ 
  & (0.08) & (0.08) & (0.06) & (0.07) & (0.09) & (0.08) \\ 
 EMI(t)*Pol(t) &  & 0.23 &  & 0.13 &  & 0.002 \\ 
  &  & (0.30) &  & (0.25) &  & (0.10) \\ 
 Intercept & $-$0.16 & $-$0.14 & $-$0.25$^{*}$ & $-$0.43$^{**}$ & 0.03 & 0.01 \\ 
  & (0.12) & (0.16) & (0.11) & (0.13) & (0.09) & (0.12) \\ 
\hline \\[-1.8ex] 
Observations & 27 & 27 & 27 & 27 & 36 & 36 \\ 
R$^{2}$ & 0.82 & 0.85 & 0.86 & 0.92 & 0.84 & 0.85 \\ 
Adjusted R$^{2}$ & 0.77 & 0.80 & 0.83 & 0.89 & 0.81 & 0.82 \\ 
Residual Std. Error & 0.47 & 0.45 & 0.42 & 0.33 & 0.43 & 0.43 \\ 
F Statistic & 18.90$^{**}$ & 15.45$^{**}$ & 25.67$^{**}$ & 30.91$^{**}$ & 31.58$^{***}$ & 23.06$^{**}$ \\ 
\hline 
\hline \\[-1.8ex] 
\textit{Note: Numbers in parentheses represent standard errors.}  & \multicolumn{6}{r}{$^{\dag}$p$<$0.1; $^{*}$p$<$0.05; $^{**}$p$<$0.01} \\ 
\end{tabular}
}
\end{table}

\begin{table}[htbp] \centering 
  \caption{Models of Congressional productivity as a function of EMI and relevant covariates. Fits start in 1879 and include the logarithm of the number of patents as an approximation of public mood in support of more or less government policies. Data end in 2004 for MLI and LPI and in 2022 for the logarithm of the number of laws passed in the session.} 
  \label{tab:Prod2} 
  \resizebox{\textwidth}{!}{
\begin{tabular}{@{\extracolsep{5pt}}lcccccc} 
\\[-1.8ex]\hline 
\hline \\[-1.8ex] 
 & \multicolumn{6}{c}{\textit{Dependent variable:}} \\ 
\cline{2-7} 
\\[-1.8ex] & \multicolumn{2}{c}{MLI} & \multicolumn{2}{c}{LPI} & \multicolumn{2}{c}{nlaw} \\ 
\\[-1.8ex] & (1) & (2) & (3) & (4) & (5) & (6)\\ 
\hline \\[-1.8ex] 
 MLI(t-1) & \textbf{0.67$^{**}$} & \textbf{0.59$^{**}$} &  &  &  &  \\ 
  & (0.09) & (0.08) &  &  &  &  \\ 
 EMI(t) &  & \textbf{0.11$^{*}$} &  & 0.07 &  & 0.05 \\ 
  &  & (0.05) &  & (0.05) &  & (0.10) \\ 
 LPI(t-1) &  &  & \textbf{0.63$^{**}$} & \textbf{0.58$^{**}$} &  &  \\ 
  &  &  & (0.11) & (0.11) &  &  \\ 
 nlaw(t-1) &  &  &  &  & \textbf{0.33$^{**}$} & \textbf{0.32$^{*}$} \\ 
  &  &  &  &  & (0.12) & (0.12) \\ 
 Pol(t) & \textbf{$-$0.19$^{**}$} & \textbf{$-$0.20$^{**}$} & \textbf{$-$0.29$^{**}$} & \textbf{$-$0.31$^{**}$} & \textbf{$-$0.43$^{**}$} & \textbf{$-$0.41$^{**}$} \\ 
  & (0.07) & (0.07) & (0.11) & (0.11) & (0.11) & (0.13) \\ 
 patents(t) & \textbf{0.28$^{*}$} & \textbf{0.37$^{**}$} & \textbf{0.18$^{*}$} & \textbf{0.22$^{*}$} & $-$0.10 & $-$0.11 \\ 
  & (0.11) & (0.10) & (0.09) & (0.08) & (0.08) & (0.12) \\ 
 PartyControl(t) & 0.07 & 0.10 & \textbf{0.14$^{*}$} & \textbf{0.17$^{*}$} & 0.23 & 0.25 \\ 
  & (0.07) & (0.07) & (0.07) & (0.07) & (0.15) & (0.16) \\ 
 PartyControlDif(t) & 0.10 & 0.11 & 0.04 & 0.04 & $-$0.25 & $-$0.25 \\ 
  & (0.08) & (0.08) & (0.07) & (0.07) & (0.17) & (0.16) \\ 
 EMI(t)*Pol(t) &  & $-$0.02 &  & 0.003 &  & $-$0.05 \\ 
  &  & (0.09) &  & (0.08) &  & (0.10) \\ 
 Intercept & $-$0.02 & $-$0.06 & $-$0.08$^{\dag}$ & $-$0.11$^{*}$ & $-$0.02 & $-$0.06 \\ 
  & (0.05) & (0.06) & (0.05) & (0.05) & (0.11) & (0.13) \\ 
\hline \\[-1.8ex] 
Observations & 62 & 62 & 62 & 62 & 70 & 70 \\ 
R$^{2}$ & 0.94 & 0.94 & 0.94 & 0.95 & 0.63 & 0.63 \\ 
Adjusted R$^{2}$ & 0.93 & 0.93 & 0.94 & 0.94 & 0.60 & 0.59 \\ 
Residual Std. Error & 0.26 & 0.25 & 0.24 & 0.24 & 0.63 & 0.64 \\ 
F Statistic & 167.22$^{**}$ & 125.51$^{**}$ & 186.41$^{**}$ & 133.02$^{**}$ & 21.39$^{**}$ & 14.98$^{**}$ \\ 
\hline 
\hline \\[-1.8ex] 
\textit{Note: Numbers in parentheses represent standard errors.}  & \multicolumn{6}{r}{$^{\dag}$p$<$0.1; $^{*}$p$<$0.05; $^{**}$p$<$0.01} \\ 
\end{tabular} 
}
\end{table}

\clearpage


\begin{thebibliography}{10}

\bibitem{higgins2021shared}
E.~T. Higgins, M.~Rossignac-Milon, G.~Echterhoff, {\it Current Directions in
  Psychological Science\/} {\bf 30}, 103 (2021).

\bibitem{kavanagh2018truth}
J.~Kavanagh, M.~D. Rich, Truth decay (2018).

\bibitem{lewandowsky2017beyond}
S.~Lewandowsky, U.~K. Ecker, J.~Cook, {\it Journal of applied research in
  memory and cognition\/} {\bf 6}, 353 (2017).

\bibitem{bennett2018disinformation}
W.~L. Bennett, S.~Livingston, {\it European journal of communication\/} {\bf
  33}, 122 (2018).

\bibitem{garrett2017epistemic}
R.~K. Garrett, B.~E. Weeks, {\it PloS one\/} {\bf 12}, e0184733 (2017).

\bibitem{lewandowsky2020willful}
S.~Lewandowsky, {\it Deliberate Ignorance: Choosing Not to Know\/} (The MIT
  Press, 2020), chap. Willful construction of ignorance: A tale of two
  ontologies.

\bibitem{cooper2023honest}
B.~Cooper, T.~R. Cohen, E.~Huppert, E.~E. Levine, W.~Fleeson, {\it Academy of
  Management Annals\/} {\bf 17}, 655 (2023).

\bibitem{lewandowsky2024liars}
S.~Lewandowsky, D.~Garcia, A.~Simchon, F.~Carrella, {\it Trends in Cognitive
  Sciences\/}  (2024).

\bibitem{NIPS2013_9aa42b31}
T.~Mikolov, I.~Sutskever, K.~Chen, G.~S. Corrado, J.~Dean, {\it Advances in
  Neural Information Processing Systems\/}, C.~Burges, L.~Bottou, M.~Welling,
  Z.~Ghahramani, K.~Weinberger, eds. (Curran Associates, Inc., 2013), vol.~26.

\bibitem{garten2018dictionaries}
J.~Garten, {\it et~al.\/}, {\it Behavior research methods\/} {\bf 50}, 344
  (2018).

\bibitem{Szostak24}
R.~Szostak, {\it Futures \& Foresight Science\/} p. e180 (2024).

\bibitem{graham2020democracy}
M.~H. Graham, M.~W. Svolik, {\it American Political Science Review\/} {\bf
  114}, 392 (2020).

\bibitem{finkel2020political}
E.~J. Finkel, {\it et~al.\/}, {\it Science\/} {\bf 370}, 533 (2020).

\bibitem{gentzkow2010drives}
M.~Gentzkow, J.~M. Shapiro, {\it Econometrica\/} {\bf 78}, 35 (2010).

\bibitem{jensen2012political}
J.~Jensen, {\it et~al.\/}, {\it Brookings Papers on Economic Activity\/} pp.
  1--81 (2012).

\bibitem{polacko2021causes}
M.~Polacko, {\it Statistics, Politics and Policy\/} {\bf 12}, 341 (2021).

\bibitem{sanchez2019economic}
{\'A}.~S{\'a}nchez-Rodr{\'\i}guez, G.~B. Willis, J.~Jetten,
  R.~Rodr{\'\i}guez-Bail{\'o}n, {\it European Journal of Social Psychology\/}
  {\bf 49}, 1114 (2019).

\bibitem{mccarty2016polarized}
N.~McCarty, K.~T. Poole, H.~Rosenthal, {\it Polarized America: The dance of
  ideology and unequal riches\/} (mit Press, 2016).

\bibitem{grant2008legislative}
J.~T. Grant, N.~J. Kelly, {\it Political Analysis\/} {\bf 16}, 303 (2008).

\bibitem{libgober2024comprehensive}
B.~Libgober, {\it Scientific Data\/} {\bf 11}, 16 (2024).

\bibitem{gennaro2022emotion}
G.~Gennaro, E.~Ash, {\it The Economic Journal\/} {\bf 132}, 1037 (2022).

\bibitem{lasser2023alternative}
J.~Lasser, {\it et~al.\/}, {\it Nature Human Behaviour\/} pp. 1--12 (2023).

\bibitem{pennebaker2003psychological}
J.~W. Pennebaker, M.~R. Mehl, K.~G. Niederhoffer, {\it Annual Review of
  Psychology\/} {\bf 54}, 547 (2003).

\bibitem{wallace1993political}
M.~D. Wallace, P.~Suedfeld, K.~Thachuk, {\it Journal of Conflict Resolution\/}
  {\bf 37}, 94 (1993).

\bibitem{pennebaker2002language}
J.~W. Pennebaker, T.~C. Lay, {\it Journal of Research in Personality\/} {\bf
  36}, 271 (2002).

\bibitem{poole2011ideology}
K.~T. Poole, H.~L. Rosenthal, {\it Ideology and congress\/}, vol.~1
  (Transaction Publishers, 2011).

\bibitem{lewis2021voteview}
B.~Lewis~Jeffrey, P.~Keith, B.~Adam, R.~Aaron, S.~Luke, Voteview: Congressional
  roll-call votes database (2021).

\bibitem{alvaredo2016distributional}
F.~Alvaredo, {\it et~al.\/}, Distributional national accounts (dina)
  guidelines: Concepts and methods used in wid. world, {\it Tech. rep.\/}, HAL
  (2016).

\bibitem{stimson2018public}
J.~Stimson, {\it Public opinion in America: Moods, cycles, and swings\/}
  (Routledge, 2018).

\bibitem{neiman2016speaking}
J.~L. Neiman, F.~J. Gonzalez, K.~Wilkinson, K.~B. Smith, J.~R. Hibbing, {\it
  Political Communication\/} {\bf 33}, 212 (2016).

\bibitem{Augoustinos2019}
M.~Augoustinos, P.~Callaghan, {\it The Social Psychology of Inequality\/}
  (Springer International Publishing, Cham, 2019), chap. The Language of Social
  Inequality, pp. 321--334.

\bibitem{habermas1984theory}
J.~Habermas, {\it The Theory of Communicative Action: Reason and the
  Rationalization of Society, Volume 1\/} (Beacon press, 1984).

\bibitem{morris2001reexamining}
J.~S. Morris, {\it Legislative Studies Quarterly\/} pp. 101--121 (2001).

\bibitem{wilson2020polarization}
A.~E. Wilson, V.~A. Parker, M.~Feinberg, {\it Current Opinion in Behavioral
  Sciences\/} {\bf 34}, 223 (2020).

\bibitem{altheide2004media}
D.~L. Altheide, {\it Political Communication,\/} {\bf 21}, 293 (2004).

\bibitem{Esser2013}
F.~Esser, {\it Mediatization as a Challenge: Media Logic Versus Political
  Logic\/} (Palgrave Macmillan UK, London, 2013), pp. 155--176.

\bibitem{hartman2022interventions}
R.~Hartman, {\it et~al.\/}, {\it Nature human behaviour\/} {\bf 6}, 1194
  (2022).

\bibitem{data_repo}
S.~T. Aroyehun, {\it et~al.\/}, Data repository for "computational analysis of
  us congressional speeches reveals a shift from evidence to intuition", \url{
  https://doi.org/10.17605/OSF.IO/Z6UTW} (2024).

\bibitem{code_repo}
S.~T. Aroyehun, {\it et~al.\/}, Code repository for "computational analysis of
  us congressional speeches reveals a shift from evidence to intuition",
  \url{https://doi.org/10.5281/zenodo.10980647} (2024).

\bibitem{gentzkow2018congressional}
M.~Gentzkow, J.~M. Shapiro, M.~Taddy, Congressional record for the 43rd-114th
  congresses: Parsed speeches and phrase counts (2018).

\bibitem{judd2017congressional}
N.~Judd, D.~Drinkard, J.~Carbaugh, L.~Young, Congressional-record: A parser for
  the congressional record (2017).

\bibitem{card2022computational}
D.~Card, {\it et~al.\/}, {\it Proceedings of the National Academy of
  Sciences\/} {\bf 119}, e2120510119 (2022).

\bibitem{dinatale2024lexpander}
A.~Di~Natale, D.~Garcia, {\it Behavior Research Methods\/} {\bf 56}, 952
  (2024).

\bibitem{rehurek_lrec}
R.~{\v R}eh{\r u}{\v r}ek, P.~Sojka, {\it {Proceedings of the LREC 2010
  Workshop on New Challenges for NLP Frameworks}\/} (ELRA, Valletta, Malta,
  2010), pp. 45--50.

\bibitem{reimers-2019-sentence-bert}
N.~Reimers, I.~Gurevych, {\it Proceedings of the 2019 Conference on Empirical
  Methods in Natural Language Processing\/} (Association for Computational
  Linguistics, 2019).

\bibitem{hosmer2013applied}
D.~W. Hosmer~Jr, S.~Lemeshow, R.~X. Sturdivant, {\it Applied logistic
  regression\/} (John Wiley \& Sons, 2013).

\bibitem{piketty2003income}
T.~Piketty, E.~Saez, {\it The Quarterly journal of economics\/} {\bf 118}, 1
  (2003).

\end{thebibliography}
\end{document}